\documentclass[12pt,sort&compress]{article}
\usepackage{graphicx}
\usepackage{bm}
\usepackage{amsmath,amssymb,color,mathrsfs}
\usepackage{footnote}
\usepackage{citesort}

\newcommand{\CJP}{Can. J. Phys. }
\newcommand{\PR}{Phys. Rev. }
\newcommand{\RMP}{Rev. Mod. Phys. }
\newcommand{\PRL}{Phys. Rev. Lett. }
\newcommand{\PRA}{Phys. Rev. A }

\newcommand{\PRE}{Phys. Rev. E }
\newcommand{\PLA}{Phys. Lett. A }

\newcommand{\SPJ}{Sov. Phys. - JETP }

\newcommand{\ZETF}{Zh. Eksp. Teor. Fiz. }
\newcommand{\ZPB}{Z. Phys. B }
\newcommand{\ZP}{Z. Phys. }

\definecolor{officegreen}{rgb}{0,0.5,0}
\definecolor{pakistangreen}{rgb}{0,0.4,0}
\definecolor{palatinatepurple}{rgb}{0.41,0.16,0.38}
\definecolor{sangria}{rgb}{0.57,0,0.04}
\definecolor{brown(traditional)}{rgb}{0.59,0.29,0}
\definecolor{bulgarianrose}{rgb}{0.28,0.02,0.03}

\begin{document}
\title{Comparative analysis of electric field influence on the quantum wells with different boundary conditions. II. Thermodynamic properties}
\author{O. Olendski\footnote{King Abdullah Institute for Nanotechnology, King Saud University, P.O. Box 2455, Riyadh 11451 Saudi Arabia; E-mail: oolendski@ksu.edu.sa}}

\maketitle

\begin{abstract}
Thermodynamic properties of the one-dimensional (1D) quantum well (QW) with miscellaneous permutations of the Dirichlet (D) and Neumann (N) boundary conditions (BCs) at its edges in the perpendicular to the surfaces electric field $\mathscr{E}$ are calculated. For the canonical ensemble, analytical expressions involving theta functions are found for the mean energy and heat capacity $c_V$ for the box with no applied voltage. Pronounced maximum accompanied by the adjacent minimum of the specific heat dependence on the temperature $T$ for the pure Neumann QW and their absence for other BCs are predicted and explained by the structure of the corresponding energy spectrum. Applied field leads to the increase of the heat capacity and formation of the new or modification of the existing extrema what is qualitatively described by the influence of the associated electric potential. A remarkable feature of the Fermi grand canonical ensemble is,  at any BC combination in zero fields, a  salient maximum of $c_V$ observed on the $T$ axis for one particle and its absence for any other number $N$ of corpuscles. Qualitative and quantitative explanation of this phenomenon employs the analysis of the chemical potential and its temperature dependence for different $N$. It is proved that critical temperature $T_{cr}$ of the Bose-Einstein (BE) condensation increases with the applied voltage for any number of particles and for any BC permutation except the ND case at small intensities $\mathscr{E}$ what is explained again by the modification by the field of the interrelated energies. It is shown that even for the temperatures smaller than $T_{cr}$ the total dipole moment $\langle P\rangle$ may become negative for the quite moderate $\mathscr{E}$. For either Fermi or BE system, the influence of the electric field on the heat capacity is shown to be suppressed with $N$ growing. Different asymptotic cases of, e.g., the small and large temperatures and low and high voltages are derived analytically and explained physically. Parallels are drawn to the similar properties of the 1D harmonic oscillator, and similarities and differences between them are discussed.
\end{abstract}

 \section{Introduction}\label{sec_Intro}
The preceding paper \cite{Olendski1} discovered, among other findings, the independence of the sign of the polarization $P_n$ on the boundary conditions (BCs) for the one-dimensional (1D) quantum well (QW) of the width $L$ placed into the uniform electric field $\mathscr{E}$ that is directed perpendicular to its confining surfaces located at $x=\pm L/2$: the polarization $P_0(\mathscr{E})$ of the ground state for any permutation of the Dirichlet (D),
\begin{equation}\label{Dirichlet1}
\Psi\left(\pm \frac{L}{2}\right)=0,
\end{equation}
and Neumann (N),
\begin{equation}\label{Neumann1}
\Psi'\left(\pm \frac{L}{2}\right)=0,
\end{equation}
edge requirements imposed on the wavefunction $\Psi(x)$ is positive for all applied voltages while its excited-state counterparts $P_n(\mathscr{E})$, $n\geq1$, for the small growing fields decrease from zero at $\mathscr{E}=0$ to the negative values, pass through the minimum and only after this start to increase crossing zero at the $n$- and BC-dependent intensity $\mathscr{E}_n^{ext}$. Immediately, one wonders: for any kind of the particles, is it possible to observe the total statistically averaged polarization that is negative at the small electric forces? Analysis below answers this question together with the thermodynamic calculations of the corresponding energy $E$ and heat capacity $c_V$. Following the previous research \cite{Olendski1}, the QW with the particular distribution of the BCs will be denoted by the two characters, where the first (second) one corresponds to the edge condition at the left (right) interface. Similar to the discussion of the spectrum $E_n(\mathscr{E})$ and polarizations $P_n(\mathscr{E})$ \cite{Olendski1}, all energies will be measured, if not specified otherwise, in units of $\pi^2\hbar^2/(2mL^2)$, which is a ground-state energy of the DD QW, while the unit of the electric field will be $\pi^2\hbar^2/(2emL^3)$, and that of the polarization  - $eL$, with $m$ being the particle mass and $e$ denoting the absolute value of the electronic charge. In addition, heat capacity is expressed below in terms of Boltzmann constant $k_B$. Discussion considers canonical as well as grand canonical ensembles. In this last case, the properties are calculated both for fermions and bosons. Also, frequently we draw parallels with the 1D harmonic oscillator (HO) with the potential (in regular units) \cite{Dalarsson1}
\begin{equation}\label{HO_Potential1}
V_{HO}(x)=\frac{1}{2}m\omega^2x^2
\end{equation}
whose energies $E_n^{HO}$, upon application of the electric voltage, are
\begin{equation}\label{HO_Energies1}
E_n^{HO}=\hbar\omega\left(n+\frac{1}{2}\right)-\frac{1}{2}\frac{e^2\mathscr{E}^2}{m\omega^2}.
\end{equation}
For this configuration, the natural units that will be used below are: for the energy, $\hbar\omega$; for the length, $x_0\equiv[\hbar/(m\omega)]^{1/2}$; for the electric field, $\hbar\omega/(ex_0)$; and for the polarization, $ex_0$.

\section{Canonical Ensemble}\label{sec_Canonical}
This type of the statistical ensemble assumes that the system under consideration is in the thermal equilibrium with the much larger bath characterized by the thermodynamic temperature $T$. The fundamental quantity here is the partition function
\begin{equation}\label{PartitionFunction1}
Z=\sum_ne^{-\beta E_n},
\end{equation}
where the summation runs over all possible quantum states, and the parameter $\beta$ is (in regular, unnormalized units) $\beta=1/(k_BT)$. The probability $w_n$ of finding particle in the state $n$ depends on the temperature and the energy $E_n$ as
\begin{equation}\label{ProbabilityCanonical1}
w_n=\frac{1}{Z}\,e^{-\beta E_n}.
\end{equation}
As a result, the mean value $\langle{\cal I}\rangle_{can}$ of any physical quantity $\cal I$ is calculated as
\begin{equation}\label{CanonicalMeanValue1}
\langle{\cal I}\rangle_{can}=\frac{1}{Z}\sum_nw_n{\cal I}_n=\frac{\sum_{n}{\cal I}_ne^{-\beta E_n}}{\sum_{n}e^{-\beta E_n}}.
\end{equation}
For the $N$ particles in the system, this equation has to be multiplied by $N$. Applying these general results to the QW with the different BCs in the electric field $\mathscr{E}$, one derives the mean values of the energy $\langle E\rangle$ and polarization $\langle P\rangle$
\begin{subequations}\label{CanonicalMeanValue2}
\begin{eqnarray}\label{CanonicalMeanEnergy2}
\langle E\rangle_{can}(\beta,\mathscr{E})=\frac{\sum_{n=0}^\infty E_ne^{-\beta E_n}}{\sum_{n=0}^\infty e^{-\beta E_n}}\\
\label{CanonicalMeanPolarization2}
\langle P\rangle_{can}(\beta,\mathscr{E})=\frac{\sum_{n=0}^\infty P_ne^{-\beta E_n}}{\sum_{n=0}^\infty e^{-\beta E_n}},
\end{eqnarray}
\end{subequations}
where in the left-hand side we have explicitly underlined that they are functions of the temperature $T$ (through the parameter $\beta$) and electric field [through the corresponding dependence of $E_n(\mathscr{E})$ and $P_n(\mathscr{E})$]. Equivalently, Eq.~\eqref{CanonicalMeanEnergy2} can be written as:
\begin{equation}\label{CanonicalMeanEnergy3}
\langle E\rangle_{can}=-\frac{\partial}{\partial\beta}\ln Z.
\end{equation}
Heat capacity at the constant volume $c_V$ is a work that has to be done to change the temperature of the system by one degree and, as a result of this, it is calculated as a derivative of the total energy with respect to the temperature $T$:
\begin{equation}\label{HeatCapacity1}
c_V=\frac{\partial}{\partial T}\langle E\rangle=-k_B\beta^2\frac{\partial}{\partial\beta}\langle E\rangle,
\end{equation}
where regular, unnormalized units have been used. Applying this generic definition to the canonical distribution from Eq.~\eqref{CanonicalMeanEnergy2}, one gets fluctuation-dissipation theorem \cite{Dalarsson1}
\begin{equation}\label{HeatCapacity2}
c_{can}(\beta,\mathscr{E})=\beta^2\left(\langle E^2\rangle_{can}-\langle E\rangle_{can}^2\right),
\end{equation}
where, for convenience of the notation, the subscript $V$ has been dropped. Energies $E_n$ and polarizations $P_n$ for the QW were calculated before \cite{Olendski1} while for the HO they are:
\begin{subequations}\label{HO_EnergyPolarization}
\begin{eqnarray}\label{HO_Energy2}
E_n^{HO}=n+\frac{1}{2}-\frac{1}{2}\mathscr{E}^2\\
\label{HO_Polarization2}
P_n^{HO}=-\frac{dE_n^{HO}}{d\mathscr{E}}=\mathscr{E}.
\end{eqnarray}
\end{subequations}
Note that, contrary to the hard-wall QW \cite{Olendski1}, for its HO counterpart the polarization is at any voltage a linear function of the field and is the same for all levels. Accordingly, its mean value for the one particle is equal to $\mathscr{E}$ too while the energy becomes:
\begin{equation}\label{HO_CanonicalEnergy1}
\left\langle E^{HO}\right\rangle_{can}=\frac{1}{2}+\frac{1}{e^\beta-1}-\frac{1}{2}\mathscr{E}^2.
\end{equation}
As a result, the electric field does {\it not} affect the HO canonical heat capacity, which reads \cite{Dalarsson1}:
\begin{equation}\label{HO_CanonicalHeatCapacity1}
c_{can}^{HO}(\beta)=\beta^2\frac{e^\beta}{(e^\beta-1)^2}.
\end{equation}
One can derive limiting cases of these dependencies:

for the small temperatures ($\beta\rightarrow\infty$):
\begin{subequations}\label{HO_CanonicalLimit1}
\begin{eqnarray}\label{HO_CanonicalLimit1_Energy}
\left\langle E^{HO}\right\rangle_{can}&=&\frac{1}{2}+e^{-\beta}+e^{-2\beta}+e^{-3\beta}+\ldots-\frac{1}{2}\mathscr{E}^2\\
\label{HO_CanonicalLimit1_HeatCapacity}
c_{can}^{HO}&=&\beta^2\left(e^{-\beta}+2e^{-2\beta}+3e^{-3\beta}+\ldots\right),
\end{eqnarray}
\end{subequations}

for the large temperatures ($\beta\rightarrow0$):
\begin{subequations}\label{HO_CanonicalLimit2}
\begin{eqnarray}\label{HO_CanonicalLimit2_Energy}
\left\langle E^{HO}\right\rangle_{can}&=&\frac{1}{\beta}+\frac{1}{12}\beta-\frac{1}{720}\beta^3+\ldots-\frac{1}{2}\mathscr{E}^2\\
\label{HO_CanonicalLimit2_Polarization}
c_{can}^{HO}&=&1-\frac{1}{12}\beta^2+\frac{1}{240}\beta^4-\ldots.
\end{eqnarray}
\end{subequations}
\begin{figure}
\centering
\includegraphics[width=\columnwidth]{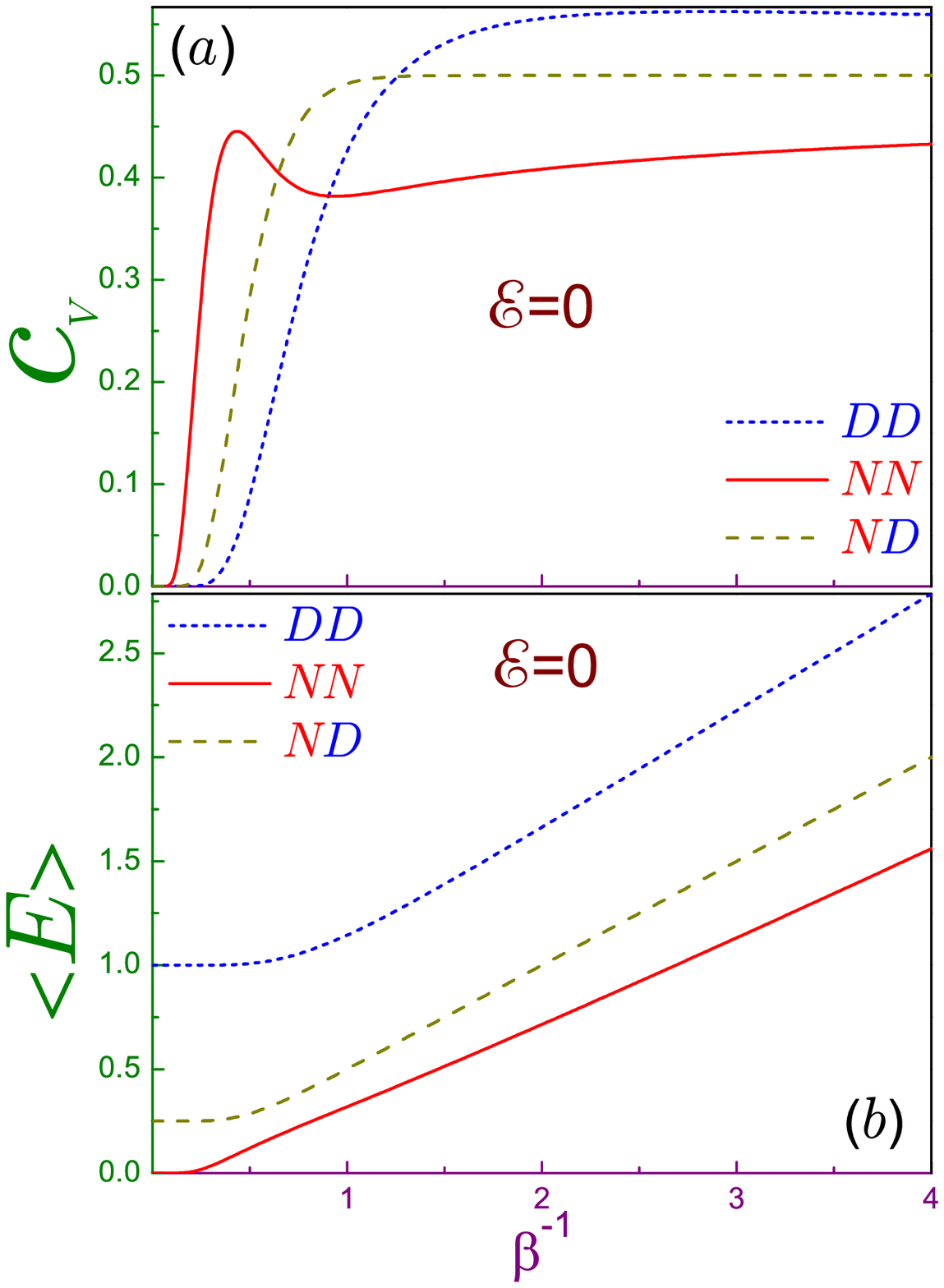}
\caption{\label{Electr0}
(a) Heat capacity $\color{pakistangreen}c_V$ and (b) mean energy $\color{pakistangreen}\langle E\rangle$ as a function of the normalized temperature $\color{palatinatepurple}\beta^{-1}$ for the canonical ensemble and pure Dirichlet (dotted line), Neumann (solid curve) and ND (dashed line) QW at zero electric field.}
\end{figure}
Before discussing the electric field influence on the thermodynamic properties of the hard-wall QW, let us address first the voltage-free configuration. Plugging in the well known expressions for the zero-field energies
\begin{equation}\label{EnergiesZeroField1}
E_n^{DD}(0)=(n+1)^2,\,E_n^{ND}(0)=\left(\!n+\frac{1}{2}\right)^2\!\!,\,E_n^{NN}(0)=n^2
\end{equation}
into Eq.~\eqref{CanonicalMeanEnergy3}, one gets after some algebra:
\begin{subequations}\label{CanonicalMeanEnergy5}
\begin{eqnarray}\label{CanonicalMeanEnergy5_DD}
\left\langle E^{DD}\right\rangle_{can}=\frac{1}{1-\theta_3\!\left(0,e^{-\beta}\right)}\frac{d}{d\beta}\,\theta_3\!\left(0,e^{-\beta}\right)\\
\label{CanonicalMeanEnergy5_ND}
\left\langle E^{ND}\right\rangle_{can}=-\frac{1}{\theta_2\!\left(0,e^{-\beta}\right)}\frac{d}{d\beta}\,\theta_2\!\left(0,e^{-\beta}\right)
\\
\label{CanonicalMeanEnergy5_NN}
\left\langle E^{NN}\right\rangle_{can}=-\frac{1}{1+\theta_3\!\left(0,e^{-\beta}\right)}\frac{d}{d\beta}\,\theta_3\!\left(0,e^{-\beta}\right).
\end{eqnarray}
\end{subequations}
Here, $\theta_i(z,q)$, $i=1,2,3,4$, are Theta functions \cite{Bellman1,Abramowitz1}. For small temperatures, $\beta\rightarrow\infty$, these equations degenerate to
\begin{subequations}\label{CanonicalMeanEnergy4}
\begin{eqnarray}\label{CanonicalMeanEnergy4_DD}
&&\left\langle E^{DD}\right\rangle_{can}=1+3e^{-3\beta}-3e^{-6\beta}+8e^{-8\beta}+3e^{-9\beta}+\ldots\\
\label{CanonicalMeanEnergy4_ND}
&&\left\langle E^{ND}\right\rangle_{can}=\frac{1}{4}+2e^{-2\beta}-2e^{-4\beta}+8e^{-6\beta}-10e^{-8\beta}+\ldots\\
&&\left\langle E^{NN}\right\rangle_{can}=e^{-\beta}-e^{-2\beta}+e^{-3\beta}+3e^{-4\beta}-4e^{-5\beta}\nonumber\\
\label{CanonicalMeanEnergy4_NN}
&&+5e^{-6\beta}-6e^{-7\beta}+3e^{-8\beta}+\ldots.
\end{eqnarray}
\end{subequations}
Utilizing transformation properties of the Theta functions \cite{Bellman1}
\begin{subequations}\label{ThetaTransformation1}
\begin{eqnarray}\label{ThetaTransformation1_1}
\theta_3\!\left(0,e^{-\beta}\right)=\sqrt{\frac{\pi}{\beta}}\,\theta_3\!\left(0,e^{-\pi^2/\beta}\right)\\
\label{ThetaTransformation1_2}
\theta_2\!\left(0,e^{-\beta}\right)=\sqrt{\frac{\pi}{\beta}}\,\theta_4\!\left(0,e^{-\pi^2/\beta}\right),
\end{eqnarray}
\end{subequations}
one derives the energies in the opposite limit of the high temperatures:
\begin{subequations}\label{CanonicalMeanEnergy6}
\begin{eqnarray}\label{CanonicalMeanEnergy6_DD}
\Bigl\langle E^{_{NN}^{DD}}\!\!\Bigr\rangle_{can}&=&\frac{1}{2\beta}\pm\frac{1}{2\pi^{1/2}\beta^{1/2}}+\frac{1}{\beta}\,e^{-\pi^2/\beta},\quad\beta\rightarrow0\\
\label{CanonicalMeanEnergy6_ND}
\left\langle E^{ND}\right\rangle_{can}&=&\frac{1}{2\beta}-\frac{2\pi^2}{\beta^2}\,e^{-\pi^2/\beta},\quad\beta\rightarrow0.
\end{eqnarray}
\end{subequations}
The corresponding heat capacities $c_V$ are calculated by applying the right-most part of Eq.~\eqref{HeatCapacity1} to the above dependencies; in particular, one has

for the ``cold" QW, $\beta\rightarrow\infty$:
\begin{subequations}\label{CanonicalHeatCapacity1}
\begin{eqnarray}\label{CanonicalHeatCapacity1_DD}
c_{can}^{DD}&=&\beta^2\left(9e^{-3\beta}-18e^{-6\beta}+64e^{-8\beta}+\ldots\right)\\
\label{CanonicalHeatCapacity1_ND}
c_{can}^{ND}&=&\beta^2\left(4e^{-2\beta}-8e^{-4\beta}+48e^{-6\beta}-80e^{-8\beta}+\ldots\right)\\
c_{can}^{NN}&=&\beta^2\left(e^{-\beta}-2e^{-2\beta}+3e^{-3\beta}+12e^{-4\beta}-20e^{-5\beta}\right.\nonumber\\
\label{CanonicalHeatCapacity1_NN}
&+&30e^{-6\beta}\left.-42e^{-7\beta}\!\!+24e^{-8\beta}\!\!+\ldots\right);
\end{eqnarray}
\end{subequations}

for the hot thermal bath:
\begin{subequations}\label{CanonicalHeatCapacity2}
\begin{eqnarray}\label{CanonicalHeatCapacity2_DD}
c_{can}^{DD,NN}=\frac{1}{2}\pm\frac{1}{4\pi^{1/2}}\beta^{1/2}-\frac{\pi^2}{\beta}\,e^{-\pi^2/\beta},\quad\beta\rightarrow0\\
\label{CanonicalHeatCapacity2_ND}
c_{can}^{ND}=\frac{1}{2}-\frac{4\pi^2}{\beta}\,e^{-\pi^2/\beta}+\frac{2\pi^4}{\beta^2}\,e^{-\pi^2/\beta},\quad\beta\rightarrow0.
\end{eqnarray}
\end{subequations}
\begin{figure}
\centering
\includegraphics[width=\columnwidth]{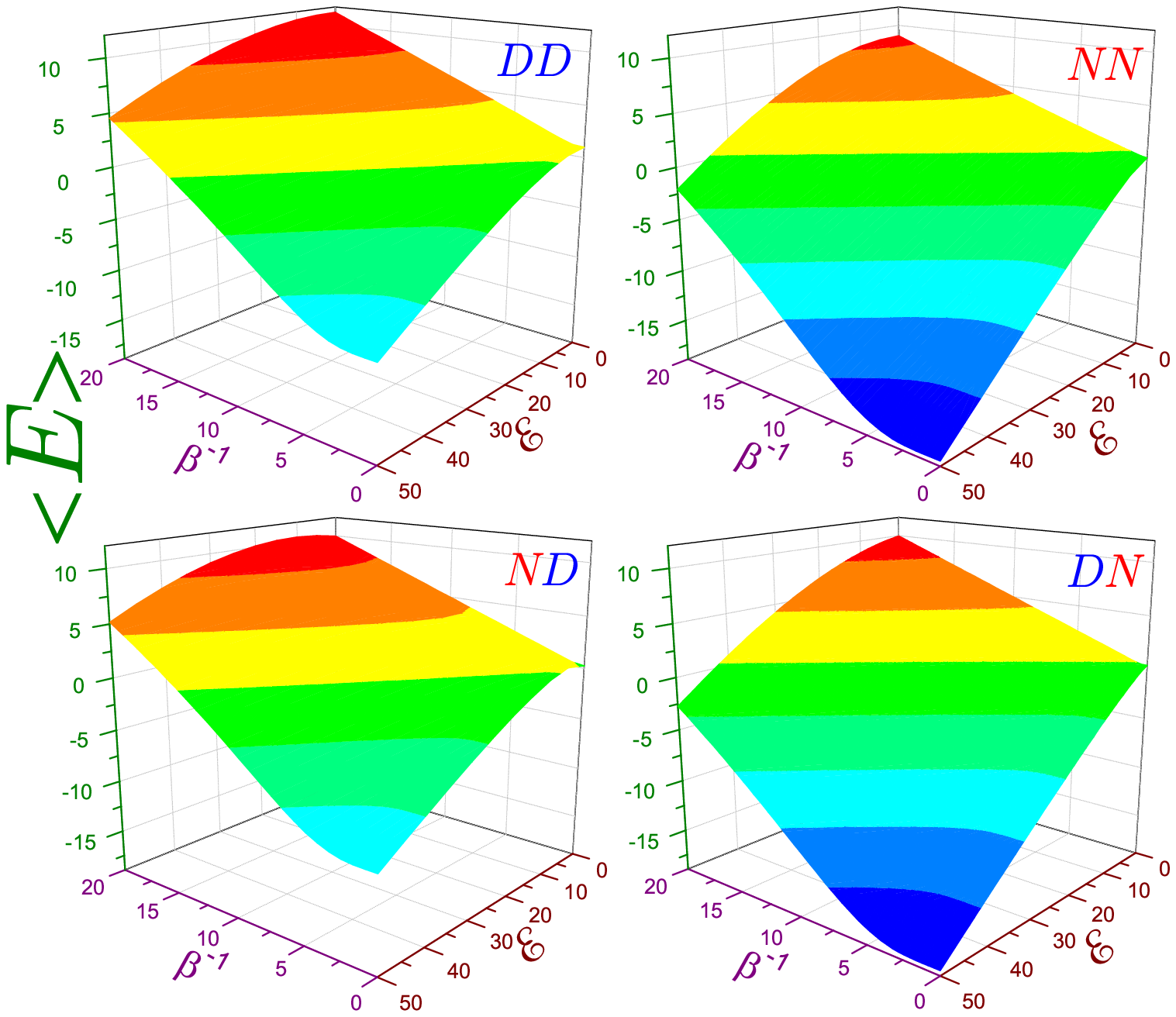}
\caption{\label{EnergyCanonical}
Mean energy $\color{pakistangreen}\langle E\rangle$ of the canonical ensemble in terms of the electric field $\color{sangria}\mathscr{E}$ and temperature $\color{palatinatepurple}\beta^{-1}$ for all permutations of the BCs. In each of the panels, the corresponding type of the edge requirements is denoted by the two characters.}
\end{figure}

Statistically averaged energies and corresponding heat capacities are shown in Fig.~\ref{Electr0}. At the zero temperature, the total internal energy reduces to the ground-state energy and the heat capacity is zero, as it should be. As it follows from Eqs.~\eqref{CanonicalMeanEnergy4} and \eqref{CanonicalHeatCapacity1}, their avalanche growth from the $T=0$ values for the purely Neumann QW takes place at the smaller temperatures as compared to the mixed BCs, which, in turn, is followed by the quantities for the Dirichlet structure. This is explained by the growing difference between the two lowest energies for just consecutively mentioned BC configurations: the quantity
\begin{equation}\label{LevelDifference1}
\Delta_n(\mathscr{E})=E_{n+1}(\mathscr{E})-E_n(\mathscr{E})
\end{equation}
at the zero field is the smallest (largest) for the Neumann (Dirichlet) structure:
\begin{equation}\label{LevelDifference2}
\Delta_n^{NN}(0)=2n+1,\quad\Delta_n^{ND}(0)=2n+2,\quad\Delta_n^{DD}(0)=2n+3.
\end{equation}

A remarkable feature of the heat capacity dependence is its nonmonotonic behaviour for the Neumann QW: at $\beta^{-1}_{max}=0.4342$ ($\beta_{max}=2.3031$) it reaches a pronounced maximum $c_{max}^{NN}=0.4455$ that is followed by the minimum of $c_{min}^{NN}=0.3818$ located at $\beta^{-1}_{min}=0.9420$ ($\beta_{min}=1.0616$). If the maximum is observed quite exactly by keeping only the  first term in the parentheses of the right-hand side of Eq.~\eqref{CanonicalHeatCapacity1_NN}, the emergence and precision of the location and magnitude of the second extremum are described better by keeping more terms in the same expansion. Physically, this nonmonotonicity of the heat capacity is attributed to the structure of the energy spectrum, see Eqs.~\eqref{EnergiesZeroField1} and \eqref{LevelDifference2}; namely, very small temperature promotes the particle mainly to the first excited level that is only one unit above the ground state, $\Delta_0^{NN}(0)=1$, with the contribution of the other levels being negligibly small due to the almost vanishing exponents in Eq.~\eqref{CanonicalMeanEnergy2} or, equivalently, in Eqs.~\eqref{CanonicalMeanEnergy4_NN} and \eqref{CanonicalHeatCapacity1_NN}; as a result, the heat capacity grows rapidly. For the larger temperatures, the occupations of the higher lying levels become essential;  however, the transitions to them are more difficult since the difference between, e.g., second and first excited states $\Delta_1^{NN}(0)=3$ is three times larger than that between the latter and the ground level. Accordingly, the same speed of the heat capacity change can not be sustained what results in the observed maximum. For the other BCs, the ratio $\Delta_1(0)/\Delta_0(0)$ is smaller than for the Neumann QW, as it follows from Eq.~\eqref{LevelDifference2}: $\Delta_1^{ND}(0)/\Delta_0^{ND}(0)=2$ and $\Delta_1^{DD}(0)/\Delta_0^{DD}(0)=5/3$; as a result, for them no extrema are observed on the $c_V-T$ dependence at $\beta^{-1}\lesssim1$. Mathematically, the drop of the NN specific heat is caused by the interplay between the counterbalancing terms $\beta^2$ and $e^{-\beta}$ in Eq.~\eqref{CanonicalHeatCapacity1_NN} as the temperature grows. Keeping only the first exponent in the parentheses of the right-hand side of this equation produces $\beta_{max}^{NN_{(1)}}=2$ while the same procedure applied to the other BCs, see Eqs.~\eqref{CanonicalHeatCapacity1_DD} and \eqref{CanonicalHeatCapacity1_ND}, results in $\beta_{max}^{DD_{(1)}}=2/3$ and $\beta_{max}^{ND_{(1)}}=1$, which are, respectively, three and two times smaller and lie beyond the range of the validity of these expansions. Accordingly, for the latter two configurations, it is essential to keep other items in the corresponding series in order for them to be correct at the decreasing $\beta$, and these extra exponents eliminate the resonance of the first-term approximation while for the Neumann QW the (negative) second component simply improves the previous result. Note that the HO leading term of the capacity expansion from Eq.~\eqref{HO_CanonicalLimit1_HeatCapacity} also results in $\beta_{max}^{HO_{(1)}}=2$; however, the subsequent (all positive) items in the series wipe out the extremum. Very broad and gentle asymmetric maximum is observed at $\beta^{-1}\gtrsim2.5$ for the Dirichlet QW while for the mixed BC the heat capacity is a monotonically increasing function of the temperature, which, at quite large $T$, rapidly approaches the asymptotic value of one half. On the contrary, the heat capacity of the symmetric QWs reaches the same limit much slower, as Eq.~\eqref{CanonicalHeatCapacity2_DD} asserts and panel (a) of Fig.~\ref{Electr0} exemplifies. Note that the HO internal energy for the high temperatures is twice of that for the hard-wall QW: $k_BT$ and $\frac{1}{2}k_BT$ in regular units, respectively. From point of view of classical equilibrium statistics that is applicable for $T\rightarrow\infty$, this difference is explained by the fact that in the former case the kinetic and potential parts of the motion make equal contributions of $\frac{1}{2}k_BT$ to the total energy \cite{Dalarsson1} while for the latter system it is the kinetic energy only that determines $\langle E\rangle$ as the QW potential is zero. As a direct consequence of this, the QW heat capacities in the same limit are one half of their HO counterpart.

\begin{figure*}
\centering
\includegraphics[width=\textwidth]{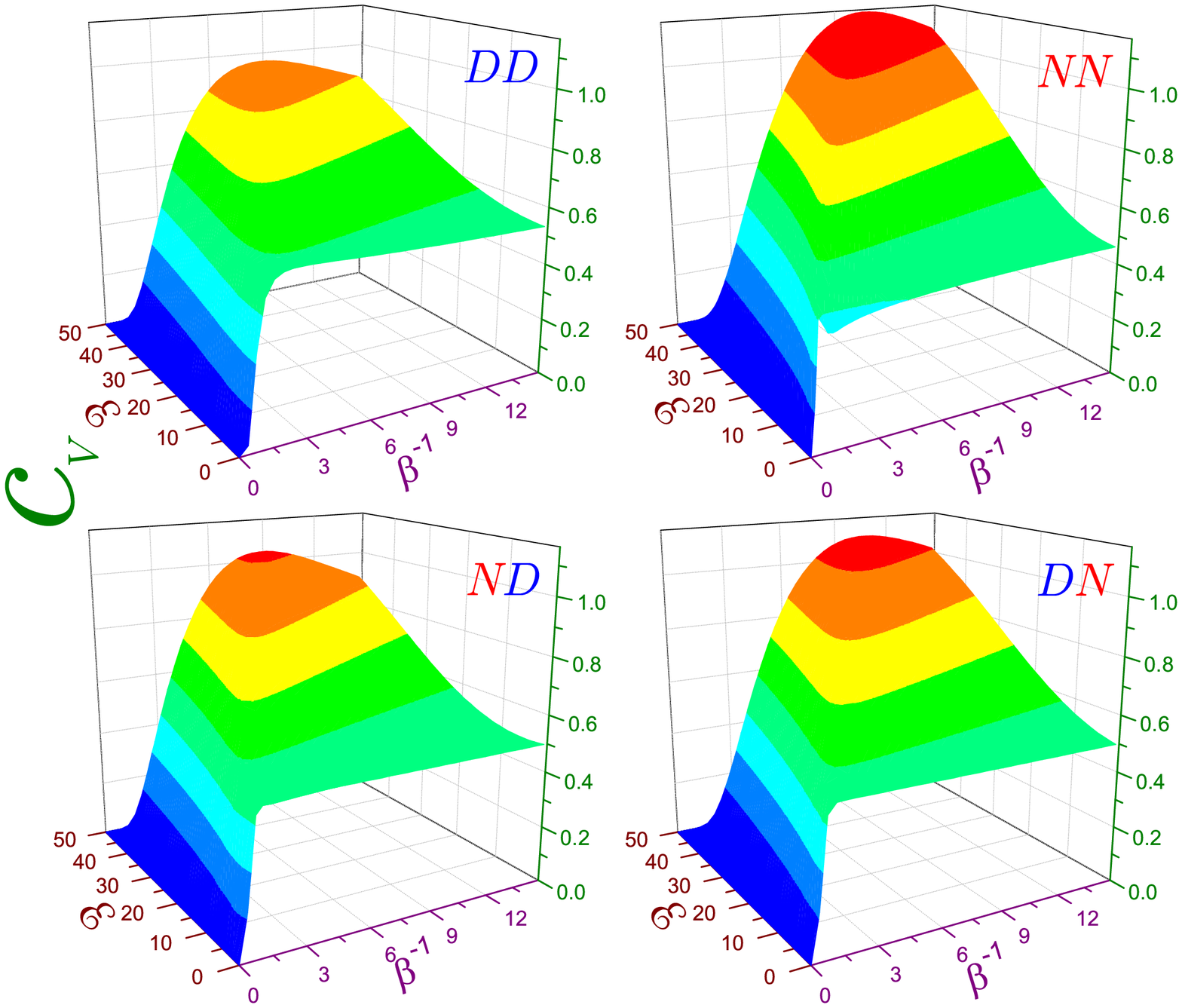}
\caption{\label{HeatCapacityCanonical}
The same as in Fig.~\ref{EnergyCanonical} but for the heat capacity $\color{pakistangreen}c_V$.}
\end{figure*}

Applied electric field modifies the energy spectrum what, in turn, affects the thermodynamic properties of the wells. It was shown that the voltage increases the difference $\Delta_0(\mathscr{E})$ between the ground and first excited levels for any permutation of the BCs (the only exception is the ND case at the small fields, see equations~(50) in  \cite{Olendski1}); accordingly, the larger temperature is needed to push out the electron from its lowest state. This is reflected in Figs.~\ref{EnergyCanonical} and \ref{HeatCapacityCanonical} where the energy $\langle E(\beta,\mathscr{E})\rangle_{can}$ and heat capacity $\langle c_V(\beta,\mathscr{E})\rangle_{can}$, respectively, are shown. It is seen that the $\beta^{-1}$ range where the mean energy does not change appreciably from the ground-state value gets wider for the stronger intensities $\mathscr{E}$. The same is true for the heat capacity where the plateau with its almost zero value grows with the field. The increasing voltage wipes out the NN minimum of the heat capacity simultaneously moving the maximum to the higher temperatures and increasing its magnitude. For each of the mixed BCs, it also creates a maximum that was absent at $\mathscr{E}=0$. Mentioned above DD extremum of the heat capacity gets narrower and its peak increases with the field growing. Recalling again the language of the classical statistical mechanics \cite{Dalarsson1}, one qualitatively explains the larger heat capacities at the nonzero fields by the contribution of the electric potential; namely, the thermally averaged value of the potential energy $\langle -\mathscr{E}x\rangle$ is: 
\begin{equation}\label{PotentialEnergy1}
\langle -\mathscr{E}x\rangle=\frac{1}{\beta}\left(1-\frac{1}{2}\beta\mathscr{E}\coth\frac{1}{2}\beta\mathscr{E}\right).
\end{equation}
This classical expression is applicable to our quantum system for the large temperatures only:
\begin{equation}\label{PotentialEnergy2}
\langle -\mathscr{E}x\rangle\approx-\frac{1}{12}\mathscr{E}^2\beta+\frac{1}{720}\mathscr{E}^4\beta^3-\ldots,\quad\beta\rightarrow0.
\end{equation}
Then, the potential contribution to the heat capacity reads:
\begin{equation}\label{HeatCapacityPotential1}
c_V^{pot}\approx\frac{1}{12}(\beta\mathscr{E})^2-\frac{1}{240}(\beta\mathscr{E})^4+\ldots,\quad\beta\rightarrow0.
\end{equation}
Note that, contrary to the HO, the kinetic and potential contributions to the heat capacity in this case, generally, are not equal to each other. Let us also mention once again that the electric field does {\it not} affect at all the HO heat capacity, see Eq.~\eqref{HO_CanonicalHeatCapacity1}, since it simply shifts all the levels by the same amount, according to Eq.~\eqref{HO_Energy2}.

\begin{figure}
\centering
\includegraphics[width=\columnwidth]{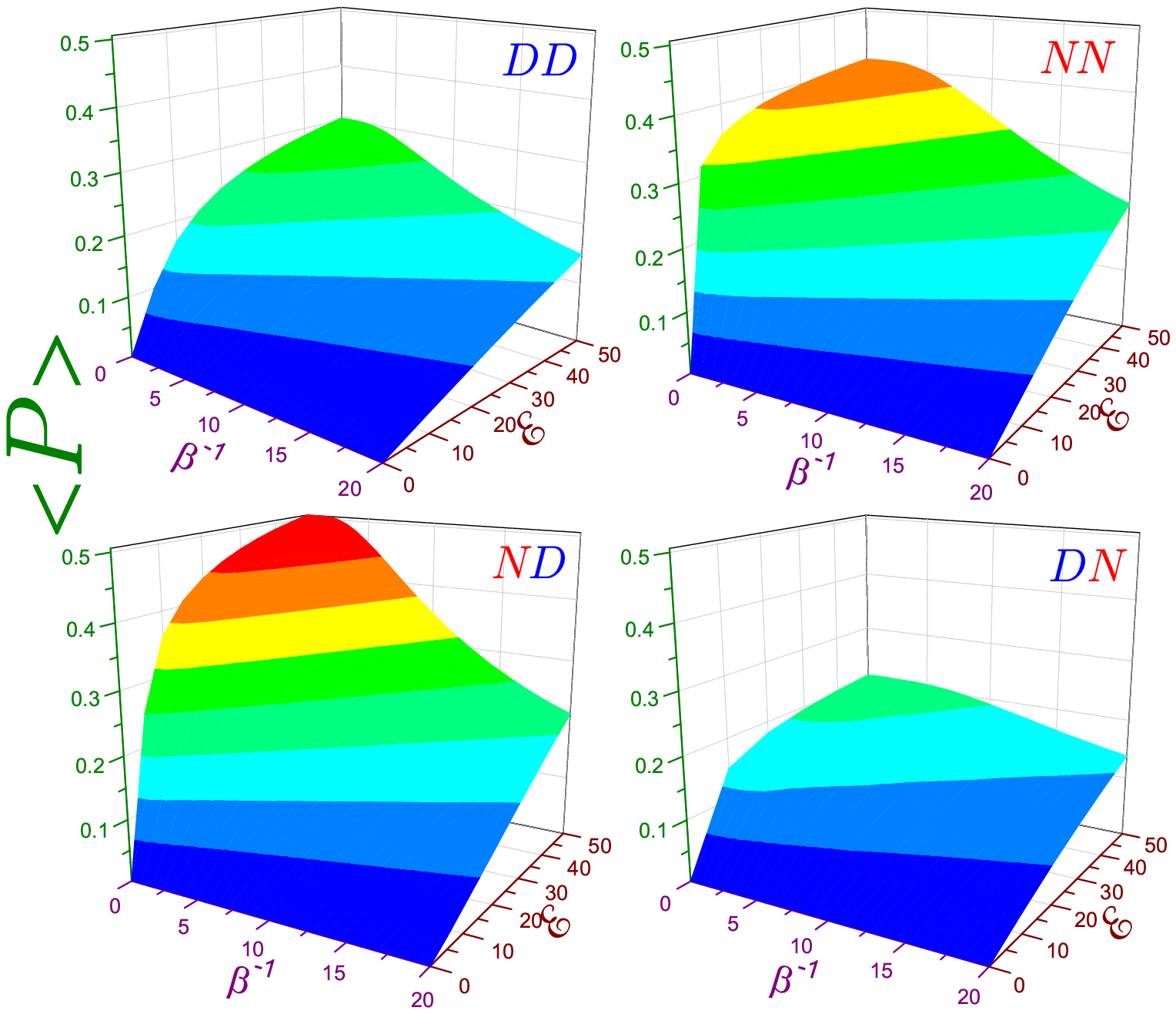}
\caption{\label{PolarizationCanonical}
The same as in Fig.~\ref{EnergyCanonical} but for the polarization $\color{pakistangreen}\langle P\rangle$.}
\end{figure}

Fig.~\ref{PolarizationCanonical} depicts statistically averaged polarizations $\langle P\rangle_{can}$ in terms of $\mathscr{E}$ and $\beta^{-1}$. Growing temperature leads to the decrease of $\langle P\rangle_{can}$ for all electric fields; however, thermal energy is not strong enough to make the total dipole moment negative: for any BC the polarization stays positive. To understand the statistical properties better, it is instructive to consider the case of the low temperatures. For the small voltages, as a first approximation, we also accept undisturbed by the field energies from Eq.~\eqref{EnergiesZeroField1}. Then, one has following dependencies: 
\begin{subequations}\label{PolarizationLimit1}
\begin{eqnarray}
\langle P^{DD}\rangle&=&P_0+\left(P_1-P_0\right)\left(e^{-3\beta}-e^{-6\beta}\right)\nonumber\\
\label{PolarizationLimit1_DD}
&+&\left(P_2-P_0\right)e^{-8\beta}+\ldots,\quad\beta\rightarrow\infty\\
\langle P^{ND}\rangle&=&P_0+\left(P_1-P_0\right)\left(e^{-2\beta}-e^{-4\beta}\right)\nonumber\\
\label{PolarizationLimit1_ND}
&+&\left(P_2-2P_0+P_1\right)e^{-6\beta}+\ldots,\quad\beta\rightarrow\infty\\
\langle P^{NN}\rangle&=&P_0+\left(P_1-P_0\right)\left(e^{-\beta}-e^{-2\beta}+e^{-3\beta}\right)\nonumber\\
\label{PolarizationLimit1_NN}
&+&\left(P_2-\!P_1\right)e^{-4\beta}+\ldots,\quad\beta\rightarrow\infty.
\end{eqnarray}
\end{subequations}
These equations were derived under the assumption of $\mathscr{E}\ll1$, but the general property stating that the first-order temperature correction is determined by $P_1-P_0$, holds for any electric intensities. For the small fields, this difference is negative \cite{Olendski1} what naturally explains the decrease of the total polarization with the temperature growing. In the opposite limit of the high voltages, the polarizations of the QW with the uniform BCs tend to the same level-independent value of one-half \cite{Olendski1} what requires larger temperatures in order to see the deviation of $\langle P\rangle$ from its $T=0$ value. This is exemplified in Fig.~\ref{PolarizationCanonical} where the temperature-independent plateau at $T=0$ widens with the field growing. For the mixed edge requirements, this limiting quantity is supplemented by the term that is proportional to $\pm(n+1/2)^{-2}$ with its sign being determined by the orientation of the BCs \cite{Olendski1}; so, for the DN case it is actually possible to observe the increase of the polarization with the temperature growing from zero. This feature is not shown in the corresponding panel of the figure since it takes place beyond the figure range $0\le\mathscr{E}\le50$.

\section{Grand Canonical Ensemble}\label{sec_GrandCanonical}
Grand canonical distribution is used for the description of the quantum system that, in addition to the thermal balance with the external reservoir, is also in the chemical  equilibrium with it. Accordingly, the structure  can exchange the energy as well as particles with the heat bath. So, the number of the quantum corpuscles $N$ in it can be changed. The fundamental role in this case is played by the chemical potential $\mu$, which is defined from the condition 
\begin{equation}\label{NumberN_1}
N=\sum_n\frac{1}{e^{(E_n-\mu)\beta}\pm1},
\end{equation}
where the upper sign corresponds to the Fermi-Dirac (FD) distribution while the lower one describes Bose-Einstein (BE) particles. Physically, the difference between these two statistics is in the fact that each quantum level can not be occupied by more than one fermion (Pauli exclusion principle) while the arbitrary number of bosons can coexist in the same state. The distribution function now depends not only on the energies $E_n$ but also on the number of the particles in the system $N$; namely, for the physical quantity $\cal I$ its grand canonical average value $\langle{\cal I}\rangle_{gc}$ is
\begin{equation}\label{GrandCanonicalMeanValue1}
\langle{\cal I}\rangle_{gc}=\sum_n\frac{{\cal I}_n}{e^{(E_n-\mu)\beta}\pm1}.
\end{equation}
Applying the distribution from Eq.~\eqref{GrandCanonicalMeanValue1} for the calculation of the heat capacity, Eq.~\eqref{HeatCapacity1}, one finds that its grand canonical value $c_{gc}$ is
\begin{equation}\label{GrandCanonicalHeatCapacity1}
c_{gc}=\beta^2\sum_n\frac{E_n\left(E_n-\mu-\beta\frac{\partial\mu}{\partial\beta}\right)}{\left[e^{(E_n-\mu)\beta}\pm1\right]^2}\,e^{(E_n-\mu)\beta},
\end{equation}
where the chemical potential, which, in the case of fermions, is also frequently called the Fermi level, is calculated, as stated above, from Eq.~\eqref{NumberN_1}. Physically, the value of $\mu$ corresponds to the energy that is needed for changing by one the number of the particles in the system:
\begin{equation}\label{ChemicalPotential1}
\mu=\left(\frac{\partial\langle E\rangle}{\partial N}\right)_{T,V}.
\end{equation}
For calculating its partial derivative with respect to the temperature, one should consider Eq.~\eqref{NumberN_1} as a condition of zeroing of the implicit function $F(\mu,\beta,N)$ of the chemical potential  in terms of the variables $\beta$ and $N$:
\begin{equation}\label{Implicit1}
F(\mu(\beta,N),\beta,N)=0.
\end{equation}
The rule of differentiating implicit functions states \cite{Rudin1}:
\begin{equation}\label{Implicit2}
\frac{\partial\mu}{\partial\beta}=-\frac{\partial F/\partial\beta}{\partial F/\partial\mu}.
\end{equation}
As a result, one finds:
\begin{equation}\label{Implicit3}
\beta\frac{\partial\mu}{\partial\beta}=\frac{\sum_n\frac{E_n-\mu}{\left[e^{(E_n-\mu)\beta}\pm1\right]^2}\,e^{(E_n-\mu)\beta}}{\sum_n\frac{1}{\left[e^{(E_n-\mu)\beta}\pm1\right]^2}\,e^{(E_n-\mu)\beta} }.
\end{equation}

The boson statistics is used for the particles with the integer spin such as photons or Cooper pairs in superconductors while the FD distribution is applied for the system of the constituents with the half integer spins; for example, the electron with its spin of $1/2$ has, for the same energy, two projections of its spin equal to $\pm1/2$. However, in our discussion below we will neglect this fact and will assume that  the number of the fermions for each energy $E_n$ is not larger than one.

\begin{figure*}
\centering
\includegraphics[width=\textwidth]{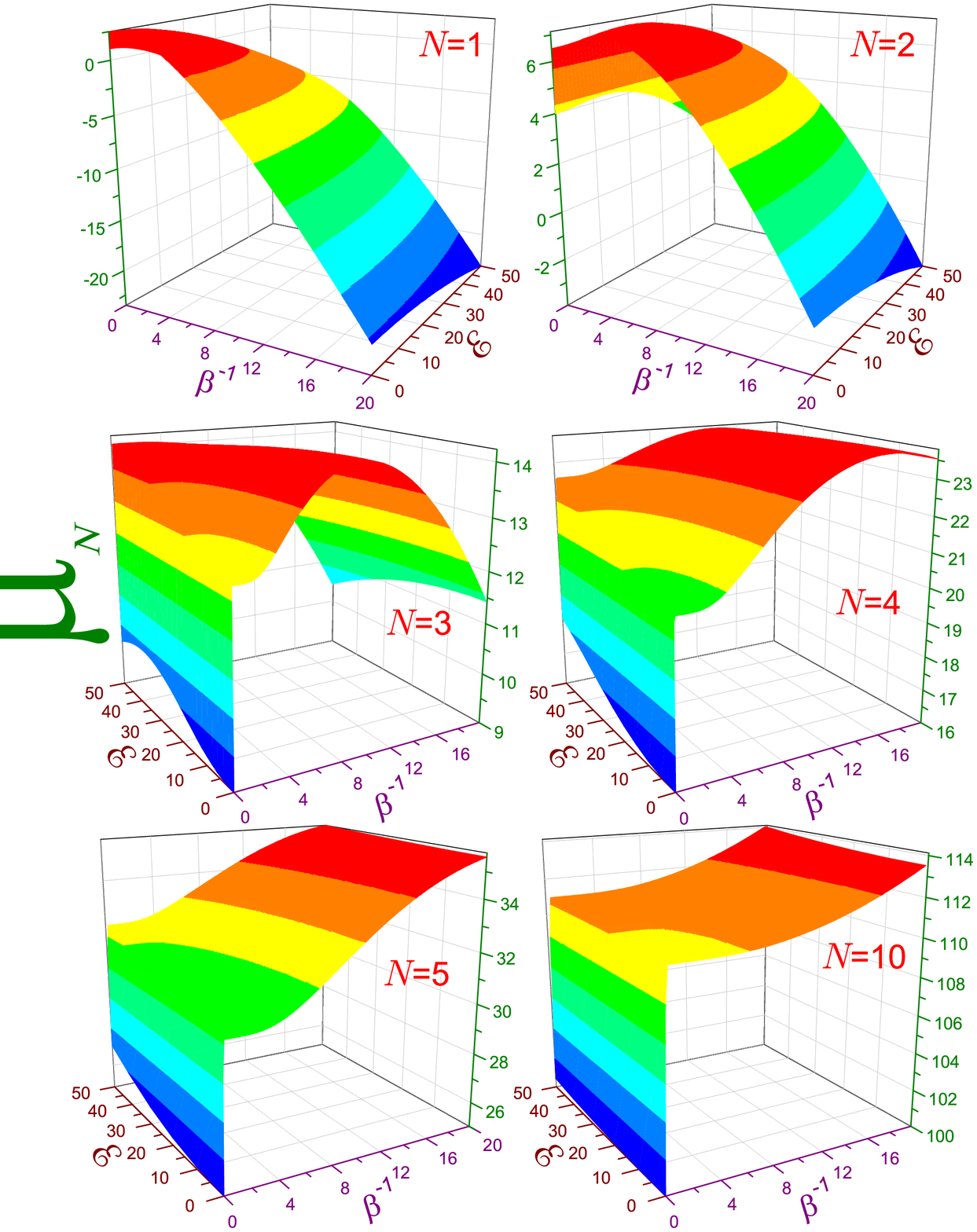}
\caption{\label{FermiGrandDirichlet}
Chemical potential $\color{pakistangreen}\mu_N$ for the FD distribution of the pure Dirichlet QW as a function of the electric field $\color{sangria}\mathscr{E}$ and temperature $\color{palatinatepurple}\beta^{-1}$ for different number of fermions $\color{red}N$ that is depicted in each panel. Note different $\color{pakistangreen}\mu$ scale for each plot. Viewing perspectives for the upper two panels are different from those for the other subplots.}
\end{figure*}

\begin{figure*}
\centering
\includegraphics[width=\textwidth]{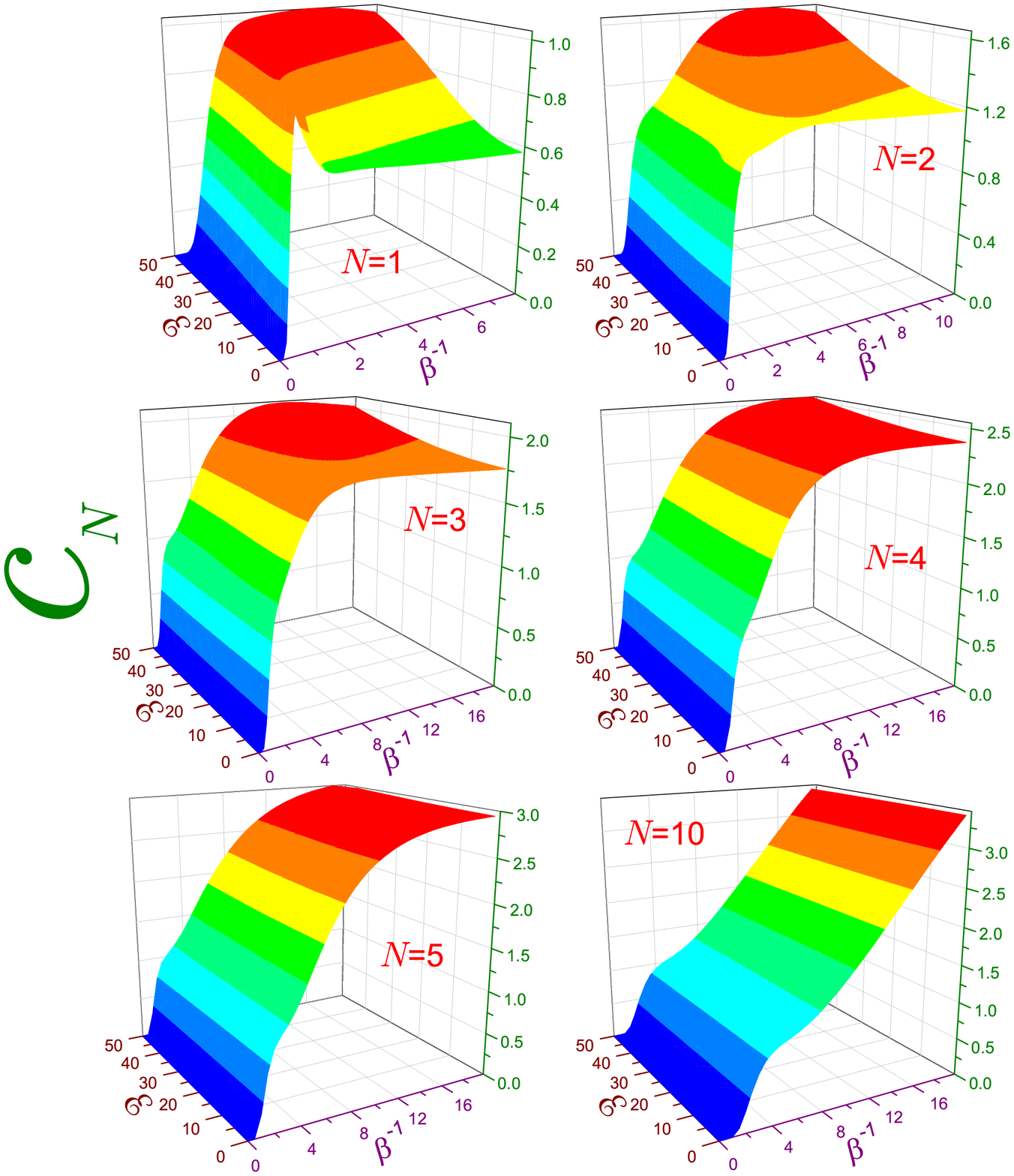}
\caption{\label{HeatCapacityGrandDirichlet}
The same as in Fig.~\ref{FermiGrandDirichlet} but for the heat capacity $\color{pakistangreen}c_N$. Note different $\color{pakistangreen}c$ and $\color{palatinatepurple}\beta^{-1}$ scales for different panels.}
\end{figure*}

Fig.~\ref{FermiGrandDirichlet} depicts the FD chemical potential for the pure Dirichlet QW as a function of the electric field and temperature for several numbers $N$. Qualitatively, the same features are characteristic for other BC permutations too. There are several distinct regions of the Fermi energy $\mu_N$ dependence on the temperature. From its $E_{N-1}$ value at $T=0$ it rapidly grows as
\begin{equation}\label{ChemicalPotentialLimit1}
\mu_N=E_{N-1}-\frac{1}{\beta}\ln\frac{1+\sqrt{1-4e^{-\Delta_{N-1}\beta}}}{2},\quad\beta\rightarrow\infty,
\end{equation}
until it reaches and stays exactly at the value of
\begin{equation}\label{ChemicalPotentialLimit2}
\mu_N=\frac{1}{2}\left(E_{N-1}+E_N\right),\quad\beta\gtrsim1,
\end{equation}
which is due to the interaction of the two corresponding levels that, at the zero temperature, were the highest occupied and lowest unoccupied states. The width of this $T$-independent plateau is determined by the number of the particles $N$ and electric field $\mathscr{E}$. For the still higher temperatures, the chemical potential for $N=1$ decreases while for the larger number of particles, $N\ge2$, it grows with $\beta^{-1}$, reaches maximum and only after that decreases, passes zero at $\beta^{(0)}$ and continues to decline into the negative part of the spectrum. For the nonpositive chemical potentials, $\mu_N\le0$, Eq.~\eqref{NumberN_1} can be cast into the form
\begin{equation}\label{NumberN_2}
\sum_{m=0}^\infty(\mp1)^me^{\mu\beta(m+1)}\sum_{n=0}^\infty e^{-\beta(m+1)E_n}=N,\quad\mu\le0.
\end{equation}
For the zero field, $\mathscr{E}=0$, the energy spectrum from Eqs.~\eqref{HO_Energy2} and \eqref{EnergiesZeroField1} simplifies this equation as follows:

for the HO:
\begin{subequations}\label{NumberN_3}
\begin{eqnarray}\label{NumberN_3_HO}
\frac{1}{2}\sum_{m=0}^\infty(\mp1)^m\frac{e^{\mu\beta(m+1)}}{\sinh\frac{1}{2}\beta(m+1)}=N,\quad\mu\le0\\
 \textnormal{and, for example, for the mixed BCs:}\nonumber\\
\label{NumberN_3_ND}
\frac{1}{2}\sum_{m=0}^\infty(\mp1)^me^{\mu\beta(m+1)}\theta_2\left(0,e^{-\beta(m+1)}\right)=N,\quad\mu\le0.
\end{eqnarray}
\end{subequations}
Putting here the chemical potential equal to zero, $\mu=0$, leads to the calculation of $\beta^{(0)}$. In known to us literature \cite{Bellman1,Abramowitz1,Gradshteyn1,Prudnikov1,Prudnikov2,Brychkov1}, there are no analytical expressions for these infinite series. However, for the very small $\beta$, the $m=0$ terms in the above equations make the most significant contributions producing the following dependencies:

for the HO:
\begin{subequations}\label{ChemicalPotentialLimit3}
\begin{eqnarray}\label{ChemicalPotentialLimit3_HO}
\left.\mu^{HO}(\beta)\right|_{\mathscr{E}=0}=\frac{1}{\beta}\ln(N\beta),\quad\beta\rightarrow0\\
 \textnormal{and, for the hard-wall QW  with the arbitrary BCs:}\nonumber\\
\label{ChemicalPotentialLimit3_ND}
\left.\mu^{IJ}(\beta)\right|_{\mathscr{E}=0}=\frac{1}{\beta}\ln\!\!\left(\!2N\sqrt{\frac{\beta}{\pi}}\right),\quad\beta\rightarrow0.
\end{eqnarray}
\end{subequations}
Superscripts $I$ and $J$ in Eq.~\eqref{ChemicalPotentialLimit3_ND} stand for any of the values of $D$ and/or $N$. Fig.~\ref{FermiGrandDirichlet} manifests that, for the larger $N$, these asymptotics are achieved at the higher temperatures. As a result, the grand canonical mean energy $\langle E\rangle_{gc}$ reads in the same limit:
\begin{subequations}\label{GrandEnergyLimit1}
\begin{eqnarray}\label{GrandEnergyLimit1_HO}
\left.\langle E^{HO}\rangle_{gc}(\beta)\right|_{\mathscr{E}=0}=\frac{N}{\beta},\quad\beta\rightarrow0\\
\label{GrandEnergyLimit1_ND}
\left.\langle E^{IJ}\rangle_{gc}(\beta)\right|_{\mathscr{E}=0}=\frac{1}{2}\frac{N}{\beta},\quad\beta\rightarrow0,
\end{eqnarray}
\end{subequations}
what, by means of Eq.~\eqref{HeatCapacity1}, immediately leads to the associated heat capacities $c_{gc}$:
\begin{subequations}\label{GrandCanonicalHeatCapacity2}
\begin{eqnarray}\label{GrandCanonicalHeatCapacity2_HO}
\left.c_{gc}^{HO}(\beta)\right|_{\mathscr{E}=0}=N,\quad\beta\rightarrow0\\
\label{GrandCanonicalHeatCapacity2_ND}
\left.c_{gc}^{IJ}(\beta)\right|_{\mathscr{E}=0}=\frac{N}{2},\quad\beta\rightarrow0.
\end{eqnarray}
\end{subequations}
A comparison of these remarkable results with Eqs.~\eqref{HO_CanonicalLimit2}, \eqref{CanonicalMeanEnergy6} and \eqref{CanonicalHeatCapacity2} confirms the general property, which states that for the large temperatures there is no difference between canonical and grand canonical distributions \cite{Dalarsson1}. However, for the small $T$ these two statistics produce very different features. Fig.~\ref{HeatCapacityGrandDirichlet} shows the FD heat capacity of the pure Dirichlet QW in terms of the temperature and electric field for the different $N$ corresponding to their counterparts from Fig.~\ref{FermiGrandDirichlet}. It is seen that, for the larger number of the particles, the asymptotics from Eq.~\eqref{GrandCanonicalHeatCapacity2_ND} is achieved at the higher $T$. At the zero field, a prominent characteristic of the heat capacity dependence for the one particle (top left panel of Fig.~\ref{HeatCapacityGrandDirichlet}) is a salient maximum $c_{max}=0.882$ observed at $\beta^{-1}_{max}=0.633$, i.e., at the right edge of the plateau from Eq.~\eqref{ChemicalPotentialLimit2}. Accordingly, we attribute this extremum to the different behaviour of the chemical potential for $N=1$  and $N\ge2$; namely, as it was mentioned during discussion of Fig.~\ref{FermiGrandDirichlet}, for one particle the Fermi energy decreases after the flat part from Eq.~\eqref{ChemicalPotentialLimit2} while for any other number $N$ it grows with $T$. Thus, their contributions to the heat capacity from Eq.~\eqref{GrandCanonicalHeatCapacity1} are opposite to each other what results in the resonance that is observed for the one particle only. Even though the shape of this maximum is quite similar to its NN counterpart for the canonical ensemble, see Sec. \ref{sec_Canonical}, its physical explanation is completely different. First, we point out that the very similar extrema are calculated also for the ND (with $c_{max}=0.879$ and $\beta^{-1}_{max}=0.418$) and pure Neumann ($c_{max}=0.878$ and $\beta^{-1}_{max}=0.208$) QWs too. The fact that the three $c_{max}$ are almost the same and the ratios of the three temperatures $T_{max}$ are practically equal to those of $\Delta_0(0)$ from Eq.~\eqref{LevelDifference2}, undoubtedly proves that the origin of this effect is the BC independent one and that the interplay between the two lowest states plays a dominant role in it. To understand these resonances, let us recall that, for the very small temperatures, the properties of the FD well are determined only by the highest occupied level and its interaction with the nearest (empty at $T=0$) above lying state, what is reflected in the extremely rapid approach by the chemical potential to the energy from Eq.~\eqref{ChemicalPotentialLimit2} that is located exactly in the middle between them. For $N\ge2$, a contribution from the lower lying members in this regime is negligibly small and can be safely neglected, while for the one-electron QW this addition is absent by definition. Further growth of the temperature increases thermal energy but it is still too ``weak'' to compel the corpuscles, which at $T=0$ lied below the Fermi energy, to contribute to the heat capacity. Only at the right edge of the plateau, the thermodynamic quantum  $k_BT$ becomes strong enough and forces other particles to donate to $\mu_N$ and $c_N$. Therefore, for $N\ge2$ the heat capacity is a quite smoothly varying function of the temperature. However, for $N=1$ there are no such additional donors that aid to support the continuous growth of the heat capacity, which can not be sustained by the one particle only. As a result, the specific heat reaches maximum and drops. 

This qualitative physical reasoning can be corroborated by the simple quantitative mathematical analysis. Fermi level from Eq.~\eqref{ChemicalPotentialLimit1} defines the corresponding mean energy and heat capacity as:
\begin{subequations}\label{GrandCanonicalLimit2}
\begin{eqnarray}\label{GrandCanonicalLimit2_Energy}
\langle E\rangle&=&\sum_{n=0}^{N-2}E_n+\frac{1}{2}E_{N-1}+E_Ne^{-\Delta_{N-1}\beta},\quad\beta\rightarrow\infty\\
\label{GrandCanonicalLimit2_HeatCapacity}
c_V&=&E_N\Delta_{N-1}\beta^2e^{-\Delta_{N-1}\beta},\quad\beta\rightarrow\infty.
\end{eqnarray}
\end{subequations}
On the other hand, for the chemical potential from Eq.~\eqref{ChemicalPotentialLimit2} these quantities for one fermion, $N=1$, become:
\begin{subequations}\label{GrandCanonicalLimit3}
\begin{eqnarray}\label{GrandCanonicalLimit3_Energy}
\langle E\rangle&=&\frac{E_0}{1+e^{-\Delta_0\beta/2}}+\frac{E_1}{1+e^{\Delta_0\beta/2}},\quad\beta\gtrsim1\\
c_V&=&\frac{1}{2}\Delta_0\beta^2\left[E_1\frac{e^{\Delta_0\beta/2}}{\left(1+e^{\Delta_0\beta/2}\right)^2}\right.\nonumber\\
\label{GrandCanonicalLimit3_HeatCapacity}
&-&\left. E_0\frac{e^{-\Delta_0\beta/2}}{\left(1+e^{-\Delta_0\beta/2}\right)^2}\right],\quad\beta\gtrsim1,
\end{eqnarray}
\end{subequations}
where we take into consideration only the two lowest states. For the pure Neumann QW without the field, this last expression takes an especially simple form:
\begin{equation}\label{GrandCanonicalLimit4_HeatCapacity}
\left.c_1^{NN}\right|_{\mathscr{E}=0}=\frac{1}{2}\beta^2\frac{e^{\beta/2}}{\left(1+e^{\beta/2}\right)^2}.
\end{equation}
This function has a pronounced maximum of $c_{max}=0.878$ at $\beta_{max}=4.799$. A perfect coincidence with the provided above exact results justifies a validity of the two-level approximation and proves that the electron transitions between them determine the specific heat resonance. It is also very instructive to contrast Eq.~\eqref{GrandCanonicalLimit4_HeatCapacity} with its  canonical counterpart from Eq.~\eqref{CanonicalHeatCapacity1_NN} in sec. \ref{sec_Canonical}. The comparison shows that the magnitude of the grand canonical Neumann extremum is almost two times larger and it is achieved at more than two times lower temperature.

Applied field $\mathscr{E}$ smooths out and widens this maximum simultaneously increasing the heat capacity. Similar to the canonical ensemble, this growth is explained by the contribution of the electrostatic potential. However, the electric influence is drastically decreased by the growing number of the particles in the QW; for example, right-bottom panels exhibit the almost full independence of the Fermi energy and heat capacity on the intensity $\mathscr{E}$ already for $N=10$. This is explained by the properties of the energy spectrum in the electric field when the higher lying states (which, in the case of the FD distribution, determine the features of the system) are less affected by the applied voltage \cite{Olendski1}.

\begin{figure}
\centering
\includegraphics[width=\columnwidth]{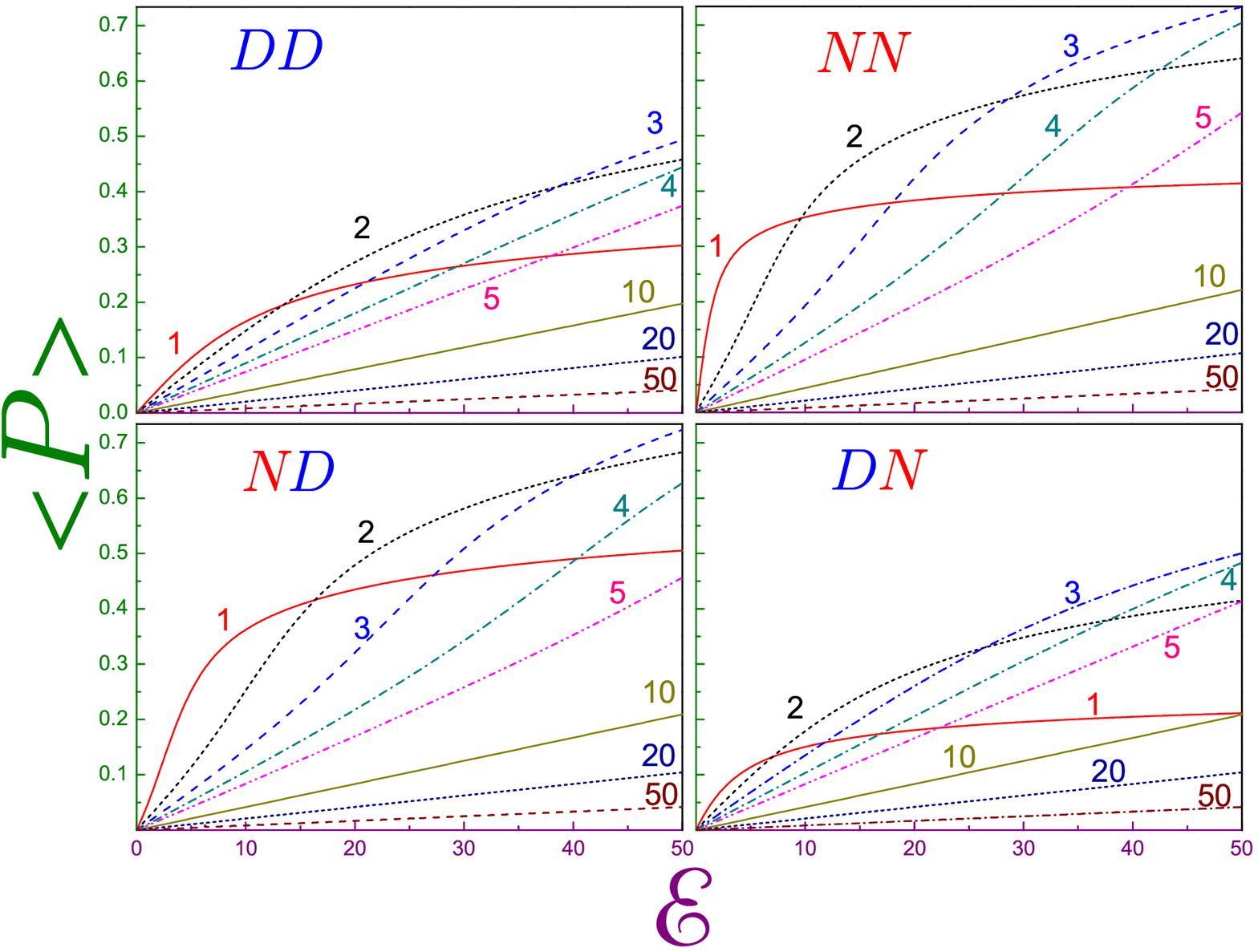}
\caption{\label{PolarizationGrandDirichletT0}
Total polarization $\color{pakistangreen}\langle P\rangle$ as a function of the electric field $\color{palatinatepurple}\mathscr{E}$ at zero temperature, $T=0$, for different number of fermions $N$ that is depicted next to the corresponding curve.}
\end{figure}

Fig.~\ref{PolarizationGrandDirichletT0} demonstrates zero-temperature FD polarization for all possible BCs and several numbers $N$. As the well accommodates more fermions, the total polarization becomes smoother function of the electric field. Figure reveals that, independently of the edge demands, the magnitude of $\langle P\rangle$ at $N\gtrsim5$ grows linearly with the voltage and the slope of this almost straight line diminishes with $N$. For any number of fermions, the total polarization remains positive at the arbitrary voltage. Nonzero temperature leads to the dependencies that qualitatively are similar to the canonical  patterns, Fig.~\ref{PolarizationCanonical}, and, because of this, the corresponding polarizations are not shown here.

Next, let us discuss bosonic structures. Remarkable experimental observations of the BE condensation in the vapors of rubidium \cite{Anderson1} and sodium \cite{Davis1} spurred an avalanche of the research on the subject predicted almost ninety years ago \cite{Bose1}, see, e. g., reviews \cite{Dalfovo1,Pitaevskii1,Leggett1,Pethick1}. Theoretically, the main effort was devoted to the calculation of the properties of the BE systems in the 3D isotropic or anisotropic harmonic traps \cite{Dalfovo1,Ketterle1,Druten1,Napolitano1,Mullin1} and their existence/nonexistence in lower dimensions \cite{Dalfovo1,Ketterle1,Druten1,Mullin1,Bagnato1}. However, other forms of the confining potentials \cite{Bagnato1,Bagnato2}, including the 3D box with the periodic \cite{Sonin1,Greenspoon1,Zasada1} or uniform \cite{Bagnato2,Goble1,Goble2,Goble3,Grossmann1} BCs, were also discussed with the comparative analysis of their influence of the properties of the trap \cite{Pathria1,Greenspoon2,Barber1,Hasan1}. From this point of view, an inclusion of the electric voltage and different BCs presents a generalization of the previous analysis. Moreover, overwhelming majority of the research concentrated on the analysis of the BE systems in the thermodynamic limit when the number of the particles and the volume containing them tend to infinity while the the density is kept constant. In this approximation, the infinite series above in this section can be safely replaced by the integrals \cite{Dalfovo1,Pitaevskii1,Pethick1}. Considering $N$ changing from one to the large values might help to understand the formation of the BE processes with the the number of the particles growing. First, we state that Eqs.~\eqref{ChemicalPotentialLimit3} - \eqref{GrandCanonicalHeatCapacity2} stay valid for the BE statistics too since they were obtained as a result of retaining the first term in the series from Eq.~\eqref{NumberN_3}. In the opposite limit of the very small temperatures, it is elementary to derive:
\begin{subequations}\label{BElimit1}
\begin{eqnarray}\label{BElimit1_ChemicalPotential}
\mu_N=E_0-\frac{1}{\beta}\ln\!\left(1+\frac{1}{N}\right),\quad\beta\rightarrow\infty,\\
 \textnormal{what leads to the mean energy and heat capacity:}\nonumber\\
\label{BElimit1_Energy}
\langle E\rangle_N=NE_0+\frac{N}{N+1}E_1e^{-\Delta_0\beta},\quad\beta\rightarrow\infty,\\
\label{BElimit1_HeatCapacity}
c_N=\frac{N}{N+1}E_1\Delta_0\beta^2e^{-\Delta_0\beta},\quad\beta\rightarrow\infty.
\end{eqnarray}
\end{subequations}
\begin{figure}
\centering
\includegraphics[width=\columnwidth]{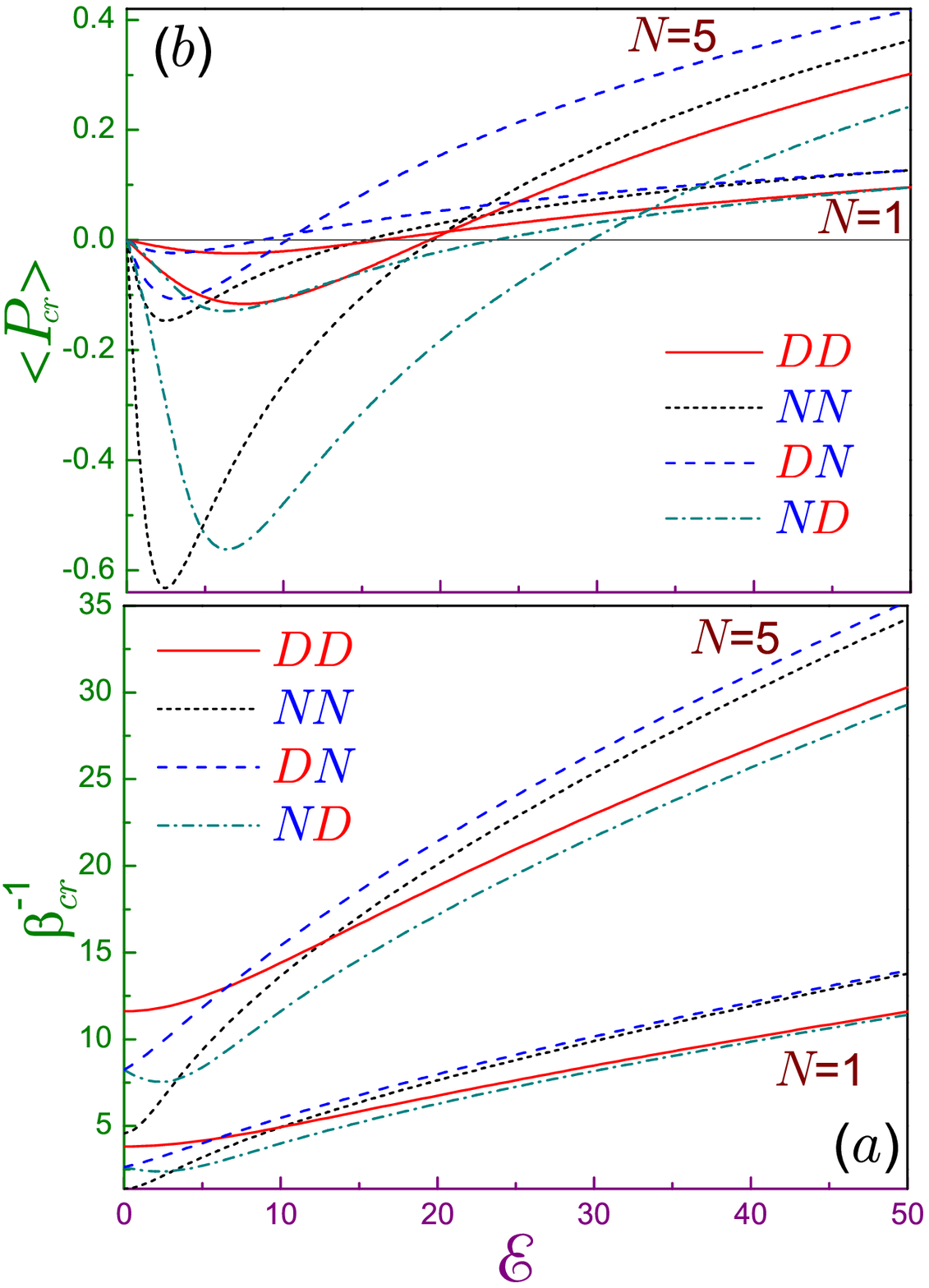}
\caption{\label{CriticalTempPolariz}
(a) Critical temperatures $\color{pakistangreen}\beta_{cr}^{-1}$ and (b) corresponding to them polarizations $\color{pakistangreen}\langle P_{cr}\rangle$ from Eq.~\eqref{FormulaCritPolariz} as functions of the applied electric field $\color{palatinatepurple}\mathscr{E}$ for different number $\color{bulgarianrose}N$ of bosons that are depicted near the corresponding curves. Solid (dotted) lines are for the pure Dirichlet (Neumann) BC while their dash (dash-dotted) counterparts denote DN (ND) geometry. In panel (b), thin horizontal line is zero polarization.}
\end{figure}

An important characteristic of the BE system is its critical temperature $T_{cr}$. It corresponds to the situation when the chemical potential is equal to the energy of the lowest level, $\mu=E_0$, and the number of the particles in this state $N_0$ is zero what leads to the implicit mathematical equation for finding $\beta_{cr}$
\begin{equation}\label{CriticalTemp1}
\sum_{n=1}\frac{1}{e^{(E_n-E_0)\beta_{cr}}-1}=N.
\end{equation}
Physically, it is the largest temperature at which the BE condensation still can be observed, and at the lower $T$ the fraction $N_0/N$ of the particles in the ground state will increase until at $T=0$ it becomes unity:
\begin{equation}\label{CriticalTemp2}
\left.N_0/N\right|_{T=0}=1.
\end{equation}

Fig.~\ref{CriticalTempPolariz}(a) shows dependencies of the critical parameter $\beta_{cr}^{-1}$ on the applied voltage for all possible BCs and several numbers $N$. In accordance with the previous results \cite{Dalfovo1}, the temperature $T_{cr}$ increases with $N$. In the absence of the fields, the lowest (highest) temperature is observed for the pure Neumann (Dirichlet) QW what is a reflection of the corresponding spectrum  from Eq.~\eqref{EnergiesZeroField1} and the energy difference between the affiliated states, see Eq.~\eqref{LevelDifference2}. Electric field leads to the modification of the mutual location of the levels on the energy axis; in particular, at the small voltages, the two lowest states move closer to each other for the ND geometry while the difference $E_1-E_0$ grows with the field for all $\mathscr{E}$ and any other BC configuration \cite{Olendski1}. As a result, the critical temperature for the former edge requirement decreases with the growing from zero field, passes through minimum and then unrestrictedly grows with the electric intensity while for all other BCs it is a continuously increasing function of $\mathscr{E}$. At the high voltages, the energy spectrum is determined mainly by the condition at the right wall \cite{Olendski1} what leads to almost the same critical temperature for, e. g., the NN and DN wells. Note that in this regime, contrary to the zero fields, the Dirichlet requirement is more favorable to the formation of the BE condensate as compared to the Neumann interface. This is explained by the larger level separation for the latter geometry at the high voltages \cite{Olendski1}.

As the ground-state polarization is positive for all fields and BCs \cite{Olendski1}, one can expect that at the onset of the BE condensation the total statistically averaged dipole moment will, at the small voltages, be negative. This is exactly what is observed in panel (b) of Fig.~\ref{CriticalTempPolariz} that shows the polarizations $\langle P_{cr}\rangle$ corresponding to the critical temperatures $\beta_{cr}^{-1}$ from panel (a). They were calculated from equation
\begin{equation}\label{FormulaCritPolariz}
\langle P_{cr}\rangle=\sum_{n=1}^\infty\frac{P_n}{e^{(E_n-E_0)\beta_{cr}}-1},
\end{equation}
and, as stated above, the critical temperature $\beta_{cr}^{-1}$ was found from Eq.~\eqref{CriticalTemp1}. The characteristic features of the critical polarizations basically follow the properties of the first excited state: from zero they decrease with the growth of the field, reach minimum after which they increase. However, for the large voltages many upper lying states are occupied and contribute to the total dipole moment. As a result, the high-field $\langle P_{cr}\rangle$ for one boson is considerably smaller than $P_1$. The absolute value of the negative polarization at the extremum grows with the number of the particles with the largest one, at the fixed $N$, being observed for the  pure Neumann QW followed by its ND counterpart what is a replica of the similar behaviour for the first excited level \cite{Olendski1}. For the temperatures below $T_{cr}$, the nonzero occupation of the ground state contributes a positive term to the polarization what leads to the gradual disappearance of the negative region of the total dipole moment $\langle P\rangle$ with the decreasing temperature until at $T=0$ it becomes $NP_0$, which is positive for any BC and arbitrary fields \cite{Olendski1}.

\begin{figure}
\centering
\includegraphics[width=\columnwidth]{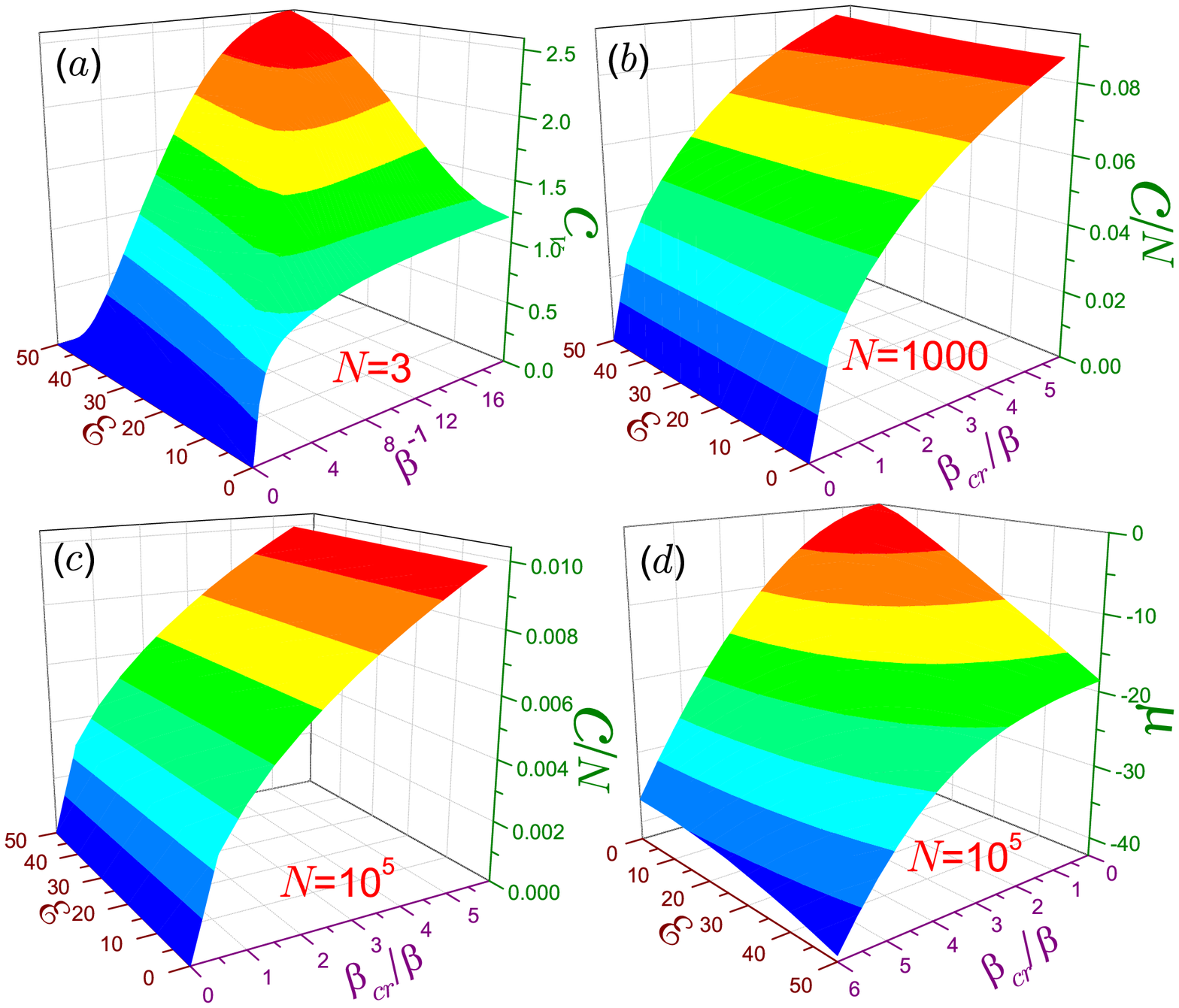}
\caption{\label{BoseNeumann}
Bosonic heat capacity $\color{pakistangreen}c_V$ of the pure Neumann QW as a function of the applied electric field $\color{sangria}\mathscr{E}$
and temperature $\color{palatinatepurple}\beta^{-1}$ for (a) $\color{red}N=3$, (b) $\color{red}N=1000$ and (c) $\color{red}N=10^5$. Note that in the last two cases the temperature is scaled in units of the critical temperature $\color{palatinatepurple}\beta_{cr}^{-1}$ while the corresponding $z$ axes measure specific heat per particle $\color{pakistangreen}c/N$. Panel (d) shows the chemical potential $\color{pakistangreen}\mu$ for $\color{red}N=10^5$ with the corresponding heat capacity depicted in part (c).}
\end{figure}

As a final example, Fig.~\ref{BoseNeumann} exhibits evolution of the heat capacity and chemical potential with the varying electric field and temperature for the pure Neumann QW. It is seen that for the small number of bosons, say, $N=3$ in panel (a), the applied voltage leads, at quite warm sample, to the increase of $c_V$ while at the small $T$, the width of the temperature-independent zero-capacity plateau increases with the field. These features were discussed before for the canonical ensemble. Increasing the number of the particles in the well leads to the suppression of the voltage dependence, as a transition from panel (a) to (b) with $N=1000$ and (c) for  $N=10^5$ vividly demonstrates. No any noticeable field dependence is seen there in the range $0\le\mathscr{E}\le50$. It is well known that for some potentials, such as, e.g., the 3D isotropic harmonic trap \cite{Pitaevskii1,Pethick1,Druten1,Napolitano1,Haugset1,Haugerud1}, the heat capacity has a cusp-like peculiarity as it passes through the critical temperature  while for the 1D quadratic potential it is a smooth function of $T$ \cite{Pethick1,Druten1,Haugset1}. Fig.~\ref{BoseNeumann} exemplifies that no any peculiarity is observed for the 1D hard-wall potential with Neumann surfaces and arbitrary applied electric fields. Our calculations confirm that the same is true for any other BCs. Finally, panel (d) shows that the chemical potential $\mu$ is a monotonically decreasing function of both the electric field $\mathscr{E}$ and temperature $T$. It is seen that the growing temperature diminishes the voltage influence on the chemical potential.

\section{Concluding remarks}\label{sec_Conclusions}
Rigorous mathematical treatment of the QW with miscellaneous BCs under the applied voltage revealed a strong influence of the interplay between them on the thermodynamic properties of the structure. In particular, without the field the differences of the energy spectrum lead, for the canonical ensemble, to the conspicuous maximum followed by the minimum of the heat capacity $c_V$ on the temperature axis for the NN quantum box while for the other edge requirements no such adjacent extrema are observed. Modification of the specific heat and statistically averaged polarization in the field is qualitatively explained by the influence of the associated electrostatic potential. Numerical calculations, which predicted, for the flat potential with the arbitrary BC, a salient maximum of $c_V$ as a function of $T$ for one fermion and its absence for the larger $N$, were corroborated by the two-level model that allows simple analytical treatment with its predictions perfectly coinciding with the exact results. From this, a clear physical explanation of this phenomenon follows that is based on the analysis of the associated Fermi energy. It is predicted that the applied field, in general, favors the formation of the BE condensate, and the differences and similarities of this process for the different BCs are discussed. The thermally averaged dipole moment is shown to take the negative values in some ranges of the fields and temperatures. It is also argued that for the larger number of either fermions or bosons in the QW, the influence of the electric field on the thermodynamic properties diminishes.

Dirichlet and Neumann conditions are the limiting cases of the so called Robin BC \cite{Gustafson1}
\begin{equation}
\left.{\bf n}{\bm\nabla}\Psi\right|_{\cal S}=\left.\frac{1}{\Lambda}\Psi\right|_{\cal S},
\end{equation}
where $\bf n$ is an inward unit normal to the surface, and the parameter $\Lambda$ has a dimension of length and is called the extrapolation distance. Its variation allows a continuous transformation from the Dirichlet ($\Lambda=0$) to the Neumann ($\Lambda=\infty$) situation. Without the field, especially intriguing are the properties of the QW at the small negative Robin lengths, $\Lambda\rightarrow-0$, when, in addition to the positive spectrum, two almost degenerate  odd and even levels with the energies $E\sim-1/(\pi\Lambda)^2$ are created \cite{Olendski2}. Analysis of the Robin QW in the electric field might present an interesting extension of the present research.

\section{Acknowledgement}\label{sec_6}
This project was supported by Deanship of Scientific Research, College of Science Research Center, King Saud University.

\end{document}